\documentclass[twocolumn,english,showpacs,superscriptaddress,prb]{revtex4}
\usepackage{graphicx}
\usepackage{amssymb}

\begin{document}

\title{Effective continuous model for surface states and thin films of
three-dimensional topological insulators}
\author{Wen-Yu Shan, Hai-Zhou Lu, and Shun-Qing Shen$^{\ast }$}
\affiliation{Centre of Theoretical and Computational Physics, The University of Hong Kong,
Pokfulam Road, Hong Kong, China}
\email{sshen@hkucc.hku.hk}

\begin{abstract}
Two-dimensional effective continuous models are derived for the surface
states and thin films of the three-dimensional topological insulator (3DTI).
Starting from an effective model for 3DTI based on the first principles
calculation [Zhang \emph{et al}, Nat. Phys. 5, 438 (2009)], we present
solutions for both the surface states in a semi-infinite boundary condition
and in thin film with finite thickness. The coupling between opposite
topological surfaces and structure inversion asymmetry (SIA) give rise to
gapped Dirac hyperbolas with Rashba-like splittings in energy spectrum.
Besides, the SIA leads to asymmetric distributions of wavefunctions for the
surface states along the film growth direction, making some branches in the
energy spectra much harder than others to be probed by light. These features
agree well with the recent angle-resolved photoemission spectra of Bi$_{2}$%
Se$_{3}$ films grown on SiC substrate [Zhang et al, arXiv:
0911.3706]. More importantly, using the parameters fitted by
experimental data, the result indicates that the thin film
Bi$_{2}$Se$_{3}$ lies in quantum spin Hall region based on the
calculation of the Chern number and the $Z_{2}$ invariant. In
addition, strong SIA always intends to destroy the quantum spin Hall
state.
\end{abstract}

\maketitle

\section{Introduction}

Topological insulators (TIs), which are band insulators with topologically
protected edge or surface states, have attracted increasing attention
recently\cite{Kane2006.Science.314.1692}. A well-known paradigm of
topological insulator is the quantum Hall effect, in which the cyclotron
motion of electrons in a strong magnetic field gives rise to insulating bulk
states but one-way conducting states propagating along edges of system\cite%
{Sarma-book}. The idea was generalized to a graphene model with spin-orbit
coupling, which exhibits the quantum spin Hall (QSH) state\cite%
{Kane2005.prl.95.226801,Kane2005.prl.95.146802}. Later, the realization of
an existing QSH matter was predicted theoretically\cite%
{Bernevig2006.science.314.1757} and soon confirmed experimentally\cite%
{Konig2007.science.318.766,Roth2009.science.325.294} in two-dimensional (2D)
HgTe/CdTe quantum wells. Furthermore it was found that the QSH\ state can be
induced even by the disorders or impurities\cite%
{Li2009.prl.102.136806,Jiang2009.prl.103.036803,Groth2009.prl.103.196805}.
Meanwhile, the concept was also generalized for three-dimensional (3D) TIs,
which are 3D band insulators surrounded by 2D conducting surface states with
quantum spin texture\cite%
{Fu2007.prl.98.106803,Moore2007.prb.75.121306,Murakami2007.njp.9.356,Teo2008.prb.78.045426}%
. Bi$_{x}$Sb$_{1-x}$, an alloy with complex structure of surface states, was
first confirmed as a 3DTI\cite%
{Hsieh2008.nature.452.970,Hsieh2009.science.323.919}. Soon after that it was
verified by both experiments\cite%
{Xia2009.natphys.5.398,Chen2009.science.325.178} and first-principles
calculations\cite{Zhang2009.natphys.5.438} that stoichiometric crystals Bi$%
_{2}$X$_{3}$ (X=Se, Te) are TIs with well-defined single Dirac cone of
surface states and extra large band gaps comparable with room temperature.
The Dirac fermions in the surface states of 3DTI obey the 2+1 Dirac
equations and reveal a lot of unconventional properties and possible
applications, such as the topological magneto-electric effect\cite%
{Qi2009.science.323.1184} and Majorana fermions for fault-tolerant quantum
computing\cite%
{Fu2008.prl.100.096407,Nilsson2008.prl.101.120403,Fu2009.prl.102.216403,Akhmerov2009.prl.102.216404,Tanaka2009.prl.103.107002,Law2009.prl.103.237001}%
.

Thanks to the state-of-art semiconductor technologies, low-dimensional
structures of Bi$_{2}$X$_{3}$ can be routinely fabricated into ultra-thin
films\cite{Zhang2009.apl.95.053314,Zhang2009.arXiv.0911.3706} and nanoribbons%
\cite{Peng2009.arXiv}. This stimulates several theoretical works on the thin
films of 3DTIs\cite{Linder2009.prb.80.205401,Lu2009.arxiv,Liu2009.arxiv}.
For further studies of the transport and optical properties of 3DTI films
and their potential applications in spintronics and quantum information, it
is desirable to establish an effective continuous model for thin films of
TIs.

In this paper, we present an effective continuous model for the surface
states and ultra-thin film of TIs. Starting with a 3D effective low-energy
model based on the first principles calculations\cite%
{Zhang2009.natphys.5.438}, we first present the solutions for the surface
states and the corresponding spectra for a semi-infinite boundary condition
of gapless Dirac Fermions and for the thin film of TIs. The finite size
effect of spatial confinement in a thin film leads to a massive Dirac model
which may exhibit the QSH\ effect. Within the same theoretical framework, a
structure inversion asymmetry (SIA) term is further introduced in this work
to account for the influence of substrate, providing a description of the
Rashba-like energy spectra observed in the angle resolved photoemission
spectra (ARPES) in the recent experiment on Bi$_{2}$Se$_{3}$ films\cite%
{Zhang2009.arXiv.0911.3706}. We derived the parameter conditions for the
formation of QSH effect in a thin film in the absence and presence of the
SIA. By analyzing the fitting parameters with the help of the Chern number
and the $Z_{2}$ invariant, we identified the ultrathin films of Bi$_{2}$Se$%
_{3}$ in the QSH phase in the experiment.

The paper is organized as follows. In Sec. \ref{sec:solutions} we introduce
an anisotropic 3D Hamiltonian for 3DTI, which is a starting point of the
present work. With this Hamiltonian, we present detailed solutions to the
thin film in two different boundary conditions. In Sec. \ref%
{sec:perturbative}, effective continuous models are established for the
surface states and thin film of 3DTI. Within the framework of this effective
continuous model, the structure inversion asymmetry is taken into account
and an effective Hamiltonian for SIA is derived in Sec. \ref{sec:SIA}. In
Sec. \ref{sec:QSH}, we apply the model to newly fabricated thin film Bi$_{2}$%
Se$_{3}$ and demonstrate that thin films of Bi$_{2}$Se$_{3}$ are in the QSH
regime. Finally, a conclusion is presented in Sec. \ref{sec:conclusion}.

\section{\label{sec:solutions}Model and general solutions for 3DTI}

\subsection{Model for 3DTI}

\begin{figure}[tbph]
\centering \includegraphics[width=0.3\textwidth]{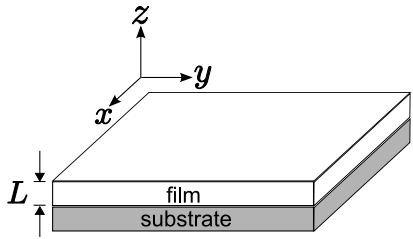}
\caption{Schematic of a topological insulator film grown on substrate. The
grown direction is defined as $z$ axis. The thickness of the film is $L$. }
\label{fig:film}
\end{figure}

As shown in Fig. \ref{fig:film}, we will consider a thin film grown along $z$
direction. The thickness of the film is $L$. We assume translational
symmetry in $x$-$y$ plane so that the wave numbers $k_{x}$ and $k_{y}$ are
good quantum numbers. We start with the effective model proposed to describe
the bulk states near the $\Gamma $ point for the bulk Bi$_{2}$Se$_{3}$\cite%
{Zhang2009.natphys.5.438}. The states are mainly contributed by four
hybridized states of Se and Bi $p_{z}$ orbitals, denoted as \{$\left\vert
p1_{z}^{+},\uparrow \right\rangle $, $\left\vert p2_{z}^{-},\uparrow
\right\rangle $, $\left\vert p1_{z}^{+},\downarrow \right\rangle $, $%
\left\vert p2_{z}^{-},\downarrow \right\rangle \}$, where $+$ ($-$) stands
for the even (odd) parity. The Hamiltonian is given by
\begin{equation}
H(\mathbf{k})=\epsilon _{0}(\mathbf{k})I_{4\times 4}+\left[
\begin{array}{cccc}
\mathcal{M}(\mathbf{k}) & -iA_{1}\partial _{z} & 0 & A_{2}k_{-} \\
-iA_{1}\partial _{z} & -\mathcal{M}(\mathbf{k}) & A_{2}k_{-} & 0 \\
0 & A_{2}k_{+} & \mathcal{M}(\mathbf{k}) & iA_{1}\partial _{z} \\
A_{2}k_{+} & 0 & iA_{1}\partial _{z} & -\mathcal{M}(\mathbf{k})%
\end{array}%
\right] ,  \label{H_zhang_natphys}
\end{equation}%
where $k_{\pm }=k_{x}\pm ik_{y}$, $\epsilon _{0}(\mathbf{k})=C-D_{1}\partial
_{z}^{2}+D_{2}k^{2}$, $\mathcal{M}(\mathbf{k})=M+B_{1}\partial
_{z}^{2}-B_{2}k^{2}$, and $k^{2}=k_{x}^{2}+k_{y}^{2}$, with $A_{1}$, $A_{2}$%
, $B_{1}$, $B_{2}$, $C$, $D_{1}$, $D_{2}$, and $M$ the model parameters.
This model has the time reversal symmetry and the inversion symmetry. Though
we start with a concrete model, the conclusion in this paper should be
applicable to other topological insulator films. We shall demonstrate that
this model for the bulk states can produce the surface states with
appropriate boundary condition.

\subsection{General solutions of the surface states}

Following the method by Zhou et al.\cite{Zhou2008.prl.101.246807}, the
general solution for either the bulk states or the surface states can be
derived analytically. Despite the existence of time-reversal symmetry, the $%
A_{2}k_{\pm }$ term couples opposite spins in Hamiltonian (1), and one has
to solve a $4\times 4$ matrix, instead of the simplified $2\times 2$ one in
the 2D case\cite{Zhou2008.prl.101.246807}. By putting a four-component trial
solution
\begin{equation}
\psi =\psi _{\lambda }e^{\lambda z}
\end{equation}%
into the Schr\"{o}dinger equation ($E$ is the eigenvalue of energy)
\begin{equation}
H(k,-i\partial _{z})\psi =E\psi ,
\end{equation}%
the secular equation
\begin{equation}
\det \left\vert H(k,-i\lambda )-E\right\vert =0
\end{equation}%
gives four solutions of $\lambda (E)$, denoted as $\beta \lambda _{\alpha
}(E)$, with $\alpha \in \{1,2\}$, $\beta \in \{+,-\}$, and%
\begin{equation}
\lambda _{\alpha }(E)=\left[ -\frac{F}{2D_{+}D_{-}}+(-1)^{\alpha -1}\frac{%
\sqrt{R}}{2D_{+}D_{-}}\right] ^{\frac{1}{2}},  \label{lambda_alpha}
\end{equation}%
where for convenience we have defined
\begin{eqnarray}
F &=&A_{1}^{2}+D_{+}(E-L_{1})+D_{-}(E-L_{2}),  \nonumber \\
R &=&F^{2}-4D_{+}D_{-}[(E-L_{1})(E-L_{2})-A_{2}^{2}k_{+}k_{-}],  \nonumber \\
D_{\pm } &=&D_{1}\pm B_{1},  \nonumber \\
L_{1} &=&C+M+(D_{2}-B_{2})k^{2},  \nonumber \\
L_{2} &=&C-M+(D_{2}+B_{2})k^{2}.
\end{eqnarray}%
Because of double degeneracy, each of the four $\beta \lambda _{\alpha }(E)$
corresponds to two linearly independent four-component vectors, found as
\begin{eqnarray}
\psi _{\alpha \beta 1} &=&\left[
\begin{array}{c}
D_{+}\lambda _{\alpha }^{2}-L_{2}+E \\
-iA_{1}(\beta \lambda _{\alpha }) \\
0 \\
A_{2}k_{+}%
\end{array}%
\right] , \\
\psi _{\alpha \beta 2} &=&\left[
\begin{array}{c}
A_{2}k_{-} \\
0 \\
iA_{1}(\beta \lambda _{\alpha }) \\
D_{-}\lambda _{\alpha }^{2}-L_{1}+E%
\end{array}%
\right] .
\end{eqnarray}%
The general solution should be a linear combination of these eight functions
\begin{equation}
\Psi (E,k,z)=\sum_{\alpha =1,2}\sum_{\beta =\pm }\sum_{\gamma =1,2}C_{\alpha
\beta \gamma }\psi _{\alpha \beta \gamma }e^{\beta \lambda _{\alpha }z},
\label{generalsolution}
\end{equation}%
with the superposition coefficients $C_{\alpha \beta \gamma }$ to be
determined by boundary conditions. In the following, we will consider two
different boundary conditions: one is semi-infinite focusing on only one
surface at $z=0$; the other includes two opposite surfaces at $z=\pm L/2$.
In both cases we assume open boundary conditions ($\Psi =0$) for the surface
states at the surfaces.

\subsection{Solutions for the surface states with semi-infinite boundary
conditions}

The surface states have a finite distribution near the boundary. For a film
thick enough that the states at opposite surfaces barely couple to each
other, we can focus on just one surface. Without loss of generality, we
study a system from $z=0$ to $+\infty $. The boundary condition is given as
\begin{equation}  \label{bound_infinite}
\Psi (z=0)=0\ \mathrm{and}\ \Psi (z\rightarrow +\infty )=0.
\end{equation}
The condition of $\Psi (z\rightarrow +\infty )=0$ requires that $\Psi $
contains only the four terms in which $\beta =-$ and the real part of $%
\lambda _{\alpha }$ is positive.

Applying the boundary conditions of Eq. (\ref{bound_infinite}) to the
general solution of Eq. (\ref{generalsolution}), the secular equation of the
nontrivial solution to the coefficients $C_{\alpha \beta \gamma }$ leads to
\begin{equation}
(\lambda _{1}+\lambda _{2})^{2}=-\frac{A_{1}^{2}}{D_{+}D_{-}},
\label{lambda1+lambda2}
\end{equation}%
which along with Eq. (\ref{lambda_alpha}) gives the dispersion of the
surface states
\begin{equation}
E_{\pm }=C+\frac{D_{1}M}{B_{1}}\pm A_{2}\sqrt{1-(\frac{D_{1}}{B_{1}})^{2}}%
k+(D_{2}-\frac{B_{2}D_{1}}{B_{1}})k^{2}.  \label{E_infinite}
\end{equation}%
Near the $\Gamma $ point, the dispersion shows a massless Dirac cone in $k$
space, with the Fermi velocity $v_{\mathrm{F}}=(A_{2}/\hbar )\sqrt{1-(\frac{%
D_{1}}{B_{1}})^{2}}$, instead of plain $A_{2}/\hbar $ as in Ref. \cite%
{Zhang2009.natphys.5.438}.

The wave functions for $E_{\pm }$ are found as
\begin{eqnarray}
\Psi _{+} &=&C_{+}^{0}\left[
\begin{array}{c}
\frac{i}{2}\sqrt{\frac{D_{+}}{B_{1}}} \\
-\frac{1}{2}\sqrt{\frac{-D_{-}}{B_{1}}} \\
-\frac{1}{2}\sqrt{\frac{D_{+}}{B_{1}}}e^{i\varphi } \\
\frac{i}{2}\sqrt{\frac{-D_{-}}{B_{1}}}e^{i\varphi }%
\end{array}%
\right] (e^{-\lambda _{2}^{+}z}-e^{-\lambda _{1}^{+}z}),  \nonumber
\label{wf_infinite} \\
\Psi _{-} &=&C_{-}^{0}\left[
\begin{array}{c}
-\frac{i}{2}\sqrt{\frac{D_{+}}{B_{1}}} \\
\frac{1}{2}\sqrt{\frac{-D_{-}}{B_{1}}} \\
-\frac{1}{2}\sqrt{\frac{D_{+}}{B_{1}}}e^{i\varphi } \\
\frac{i}{2}\sqrt{\frac{-D_{-}}{B_{1}}}e^{i\varphi }%
\end{array}%
\right] (e^{-\lambda _{2}^{-}z}-e^{-\lambda _{1}^{-}z}),
\end{eqnarray}%
where $\lambda _{\alpha }^{\pm }$ are short for $\lambda _{\alpha }(E=E_{\pm
})$ according to Eq. (\ref{lambda_alpha}), $\tan \varphi \equiv k_{y}/k_{x}$%
, and $C_{\pm }^{0}$ are the normalization factors. The properties of the
solution to $\lambda _{\alpha }$ determine the spatial distribution of the
wave functions. Generally speaking, the edge states exist if $\lambda _{1}$
and $\lambda _{2}$ are both real or complex conjugate partners. For either
case, there should be inequality relations
\begin{eqnarray}
\frac{M}{B_{1}} &>&0,  \nonumber \\
D_{+}D_{-} &<&0\ .
\end{eqnarray}%
The edge states distribute mostly near the surface ($z=0$), with the scale
of the decay length about $\lambda _{1,2}^{-1}$ for real $\lambda _{1,2}$ or
$[\mathrm{Re}(\lambda _{1,2})]^{-1}$ for complex $\lambda _{1,2}$. In the
former case, the wavefunctions decay exponentially and monotonously away
from the surface (not from $z=0$); while in latter case the decaying is
accompanied by a periodical oscillation, which can be easily seen from the
wavefunctions in Eq. (\ref{wf_infinite}). In addition, there exist complex
solutions to $\lambda _{\alpha }$ when
\begin{equation}
\frac{A_{1}^{2}}{-D_{+}D_{-}}<\frac{4M}{B_{1}}.  \label{complexlambda12}
\end{equation}

\subsection{Solutions for finite-thickness boundary conditions}

When the thickness of the film is comparable with the characteristic length $%
1/\lambda _{1,2}$ of the surface states, there is coupling between the
states on opposite surfaces. One has to consider the boundary conditions at
both surfaces simultaneously. Without loss of generality, we will consider
the top surface is located at $z=L/2$ and the bottom surface at $-L/2$. The
boundary conditions are given as
\begin{equation}
\Psi (z=\pm \frac{L}{2})=0.  \label{bound_finite}
\end{equation}%
In this case, the general solution consists of all eight linearly
independent functions. Applying the boundary conditions in Eq. (\ref%
{bound_finite}) to the general solution of Eq. (\ref{generalsolution}), the
secular equation of the nontrivial solution to the superposition
coefficients $C_{\alpha \beta \gamma }$ leads to a transcendental equation
\begin{equation}
\frac{\frac{D_{+}D_{-}}{A_{1}^{2}}(\lambda _{1}^{2}-\lambda
_{2}^{2})^{2}+(\lambda _{1}^{2}+\lambda _{2}^{2})}{\lambda _{1}\lambda _{2}}=%
\frac{\tanh \frac{\lambda _{2}L}{2}}{\tanh \frac{\lambda _{1}L}{2}}+\frac{%
\tanh \frac{\lambda _{1}L}{2}}{\tanh \frac{\lambda _{2}L}{2}}.
\label{trans_eqn}
\end{equation}%
In a large $L$ limit, $\tanh \frac{\lambda _{\alpha }L}{2}$ reduces to $1$ ,
then Eq. (\ref{trans_eqn}) can recover the result in Eq. (\ref%
{lambda1+lambda2}). With the help of Eq. (\ref{lambda_alpha}), Eq. (\ref%
{trans_eqn}) can be used to identify the energy spectra and the values of $%
\lambda _{\alpha }$ numerically.

\begin{figure}[htbp]
\centering \includegraphics[width=0.5\textwidth]{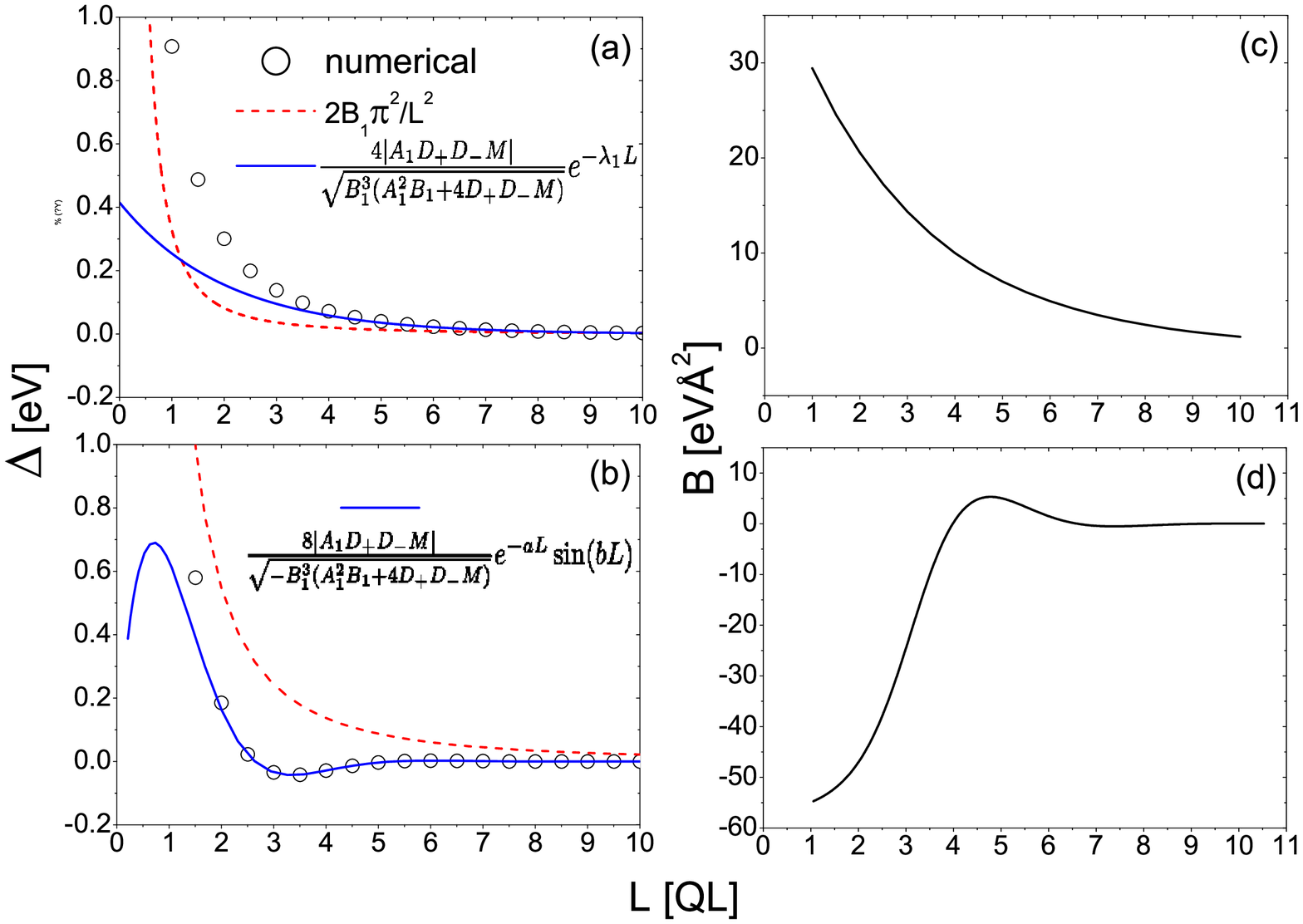}
\caption{ (Color online) [(a)(b)] The energy gap $\Delta\equiv E_+-E_-$ and
[(c)(d)] the model parameter $B$ [defined in Eq. (\protect\ref%
{paradefinition})] as functions of the film thickness $L$. Circles
correspond to the numerical results of the transcendental equations
(\protect\ref{trans_eqn_1}) and (\protect\ref{trans_eqn_2}). Solid
and dash lines correspond to the approximate formulas to $\Delta$
when $L$ is finite
[Eqs. (\protect\ref{gap_approx_real}) and (\protect\ref{gap_approx_complex}%
)] or very small [Eq. (\protect\ref{ultrathin})], respectively. All the
parameters are adopted (a) by fitting experimental results of 4QL Bi$_2$Se$%
_3 $, and (b) from the numerical fitting for the first principles
calculation of Bi$_2$Se$_3$\protect\cite{Zhang2009.natphys.5.438}, as listed
in Tab. \protect\ref{tab:4QLDiracParameters}.}
\label{fig:gap}
\end{figure}

Due to the finite size effect\cite{Zhou2008.prl.101.246807}, the coupling
between the states at the top and bottom surfaces will open an energy gap%
\cite{Lu2009.arxiv,Liu2009.arxiv,Linder2009.prb.80.205401}. We define the
gap as $\Delta =E_{+}-E_{-}$ at the $\Gamma $ point, where $E_+$ and $E_-$
are two solutions of Eq. (\ref{trans_eqn}). For $\lambda _{\alpha }L\gg 1$
and $\lambda _{2}\gg \lambda _{1}$ ($L$ can be finite), the approximate
expression for $\Delta $ can be found. If $\lambda _{\alpha }$ is real, the
gap can be approximated by
\begin{equation}  \label{gap_approx_real}
\Delta \simeq \frac{4|A_{1}D_{+}D_{-}M|}{\sqrt{
B_{1}^{3}(A_{1}^{2}B_{1}+4D_{+}D_{-}M)}}e^{-\lambda _{1}L},
\end{equation}
which decays exponentially as a function of $L$. Fig. \ref{fig:gap}(a) shows
the gap as a function of thickness, in which a set of model parameters used
to fit the ARPES of 4QL Bi$_{2}$Se$_{3}$ thin film is employed, as listed in
the first row of Tab. \ref{tab:4QLDiracParameters}.

\begin{table}[tbph]
\caption{Two sets of parameters for the 3D Dirac model. The first row is
extracted from our effective model parameters for 4QL Bi$_2$Se$_3$ film in
table \protect\ref{tab:thickness}, and the second row is adopted from the
first principles calculation\protect\cite{Zhang2009.natphys.5.438}.}
\label{tab:4QLDiracParameters}
\begin{ruledtabular}
\begin{tabular}{cccccccc}
 $M$ & $A_1$ & $A_2$ & $B_1$ & $B_2$ & $C$ & $D_1$ & $D_2$  \\ \hline
(eV) & (eV\AA ) & (eV\AA ) & (eV\AA $^2$) & (eV\AA $^2$) & (eV) & (eV\AA $^2$%
) & (eV\AA $^2$)   \\ \hline
0.28 & 3.3 & 4.1 & 1.5 & -54.1 & -0.0068 & 1.2 & -30.1 \\ \hline
0.28 & 2.2 & 4.1 & 10 & 56.6 & -0.0068 & 1.3 & 19.6  \\
\end{tabular}
\end{ruledtabular}
\end{table}

For some other materials there may exist complex $\lambda _{1}=\lambda
_{2}^{\ast }$ and we can define $\lambda _{1}=a-ib$ and $\lambda _{2}=a+ib$,
where $a>0$, $b>0$ according to Eq. (\ref{lambda_alpha}). In this case, the
gap is found as
\begin{equation}  \label{gap_approx_complex}
\Delta \simeq \frac{8|A_{1}D_{+}D_{-}M|}{\sqrt{
-B_{1}^{3}(A_{1}^{2}B_{1}+4D_{+}D_{-}M)}}e^{-aL}\sin (bL),
\end{equation}
with
\begin{eqnarray}
a &\backsimeq &\frac{A_{1}}{2\sqrt{-D_{+}D_{-}}}, \\
b &\backsimeq &\sqrt{\frac{M}{B_{1}}+\frac{A_{1}^{2}}{4D_{+}D_{-}}}.
\end{eqnarray}
According to this result, the oscillation period of the gap $\pi /b$ becomes
$\pi \sqrt{B_{1}/M}$ when $A_{1}=0$, in accordance with the result obtained
by Liu \emph{et al}\cite{Liu2009.arxiv}. Fig. \ref{fig:gap}(b) shows the gap
oscillation by using the model parameters listed in the second entry of Tab. %
\ref{tab:4QLDiracParameters}. Besides, the sine function implies that $%
\Delta $ may be negative. Later we will see that the sign of $\Delta $ can
be found by solving $E_{+}^{0}$ and $E_{-}^{0}$ from Eqs. (\ref{trans_eqn_1}%
) and (\ref{trans_eqn_2}), respectively.

\section{\label{sec:perturbative} Effective continuous models}

The solutions of the surface states and thin film of 3DTI can be
applied to calculate physical properties explicitly. For instance,
we can see whether the ground state of a thin film exhibits QSHE or
not by calculating the Chern number or Z$_{2}$ invariant. It is also
desirable to establish an effective continuous model to explore the
properties of these surface states especially when other
interactions have to be taken into account. For this purpose, in
this section we derive an effective low-energy and continuous models
for the surface states and thin film of 3DTI.

Due to the low-energy long-wavelength nature of the Dirac cone of the
surface electrons, we can use the solutions of the surface states at the $%
\Gamma $ point as a basis to expand the Hamiltonian $H(k)$ in Eq.(1), which
will be valid when the energy is limited within the band gap between the
conduction and valence bands. This is equivalent to a truncation
approximation as we exclude the solutions for the bulk states in the basis.
In this approach, the Hamiltonian in Eq. (\ref{H_zhang_natphys}) can be
expressed as
\begin{equation}
H(\vec{k})=H_{0}(k=0)+\Delta H,  \label{H_perturb}
\end{equation}%
where
\begin{equation}
H_{0}=\left[
\begin{array}{cc}
h(A_{1}) & 0 \\
0 & h(-A_{1})%
\end{array}%
\right] ,  \label{H_0}
\end{equation}%
with
\begin{equation}
h(A_{1})=\left[
\begin{array}{cc}
-D_{-}\partial _{z}^{2}+C+M & -iA_{1}\partial _{z} \\
-iA_{1}\partial _{z} & -D_{+}\partial _{z}^{2}+C-M%
\end{array}%
\right] ,
\end{equation}%
and
\begin{equation}
\Delta H=\left[
\begin{array}{cc}
D_{2}k^{2}-B_{2}k^{2}\sigma _{z} & A_{2}k_{-}\sigma _{x} \\
A_{2}k_{+}\sigma _{x} & D_{2}k^{2}-B_{2}k^{2}\sigma _{z}%
\end{array}%
\right] .
\end{equation}%
The first term can be solved exactly, and the last term describes
the behaviors of electrons near the $\Gamma $ point.

\subsection{\label{sec: Gamma point}Basis states at $\Gamma $ point}

$H_{0}$ in Eq. (\ref{H_0}) is block-diagonal. Its solution can be found by
solving each block separately, i.e., $h(A_{1})\Psi _{\uparrow }=E\Psi
_{\uparrow }$ and $h(-A_{1})\Psi _{\downarrow }=E\Psi _{\downarrow }$.
Because the lower block is the "time" reversal of the upper block, the
solutions satisfy $\Psi _{\downarrow }(z)=\Theta \Psi _{\uparrow }(z)$,
where $\Theta =-i\sigma _{y}\mathcal{K}$ is the time-reversal operator, with
$\sigma _{y}$ the $y$ component of the Pauli matrices and $\mathcal{K}$ the
complex conjugation operation. Equivalently, we can replace $A_{1}$ by $%
-A_{1}$ in all the results for the upper block, to obtain those for the
lower block. Therefore, we only need to solve $h(A_{1})$. Following the same
approach in Sec. \ref{sec:solutions}, we put a two-component trial solution
\begin{equation}
\psi ^{\uparrow }=\psi _{\lambda }^{\uparrow }e^{\lambda z}
\end{equation}%
into
\begin{equation}
h(A_{1},-i\partial _{z})\psi ^{\uparrow }=E\psi ^{\uparrow },
\end{equation}%
the secular equation for a nontrivial solution yields four roots of $\lambda
(E)$, denoted as $\beta \lambda _{\alpha }$, with $\beta \in \{+,-\}$ and $%
a\in \{1,2\}$. Note that here $\lambda _{\alpha }$ is short for $\lambda
_{\alpha }(k=0)$ in Eq. (\ref{lambda_alpha} ). Each $\beta \lambda _{\alpha
} $ corresponds to a two-component vector
\begin{equation}
\psi _{\alpha \beta }^{\uparrow }=\left[
\begin{array}{c}
D_{+}\lambda _{\alpha }^{2}-l_{2}+E \\
-iA_{1}(\beta \lambda _{\alpha })%
\end{array}%
\right] .
\end{equation}%
The general solution is a linear combination of the four linearly
independent two-component vectors
\begin{equation}
\Psi _{\uparrow }=\sum_{\alpha =1,2}\sum_{\beta =+,-}C_{\alpha \beta }\psi
_{\alpha \beta }^{\uparrow }e^{\beta \lambda _{\alpha }z}.
\label{generalsolution_2by2}
\end{equation}%
Applying the boundary conditions Eq. (\ref{bound_finite}) to this general
solution, we obtain two transcendental equations,
\begin{equation}
\frac{(C-M-E-D_{+}\lambda _{1}^{2})\lambda _{2}}{(C-M-E-D_{+}\lambda
_{2}^{2})\lambda _{1}}=\frac{\tanh (\frac{\lambda _{2}L}{2})}{\tanh (\frac{%
\lambda _{1}L}{2})},  \label{trans_eqn_1}
\end{equation}%
and
\begin{equation}
\frac{(C-M-E-D_{+}\lambda _{1}^{2})\lambda _{2}}{(C-M-E-D_{+}\lambda
_{2}^{2})\lambda _{1}}=\frac{\tanh (\frac{\lambda _{1}L}{2})}{\tanh (\frac{%
\lambda _{2}L}{2})}.  \label{trans_eqn_2}
\end{equation}%
The solutions to Eqs. (\ref{trans_eqn_1}) and (\ref{trans_eqn_2}) give two
energies at the $\Gamma $ point, designated as $E_{+}^{0}\equiv E_{+}(k=0)$
and $E_{-}^{0}\equiv E_{-}(k=0)$, respectively. The eigen wavefunctions for $%
E_{+}^{0}$ and $E_{-}^{0}$ are, respectively,
\begin{eqnarray}
\varphi (A_{1}) &\equiv &\Psi _{\uparrow }^{+}=C_{+}\left[
\begin{array}{c}
-D_{+}\eta _{1}^{+}f_{-}^{+} \\
iA_{1}f_{+}^{+}%
\end{array}%
\right] ,  \label{phichi1} \\
\chi (A_{1}) &\equiv &\Psi _{\uparrow }^{-}=C_{-}\left[
\begin{array}{c}
-D_{+}\eta _{2}^{-}f_{+}^{-} \\
iA_{1}f_{-}^{-}%
\end{array}%
\right] ,  \label{phichi2}
\end{eqnarray}%
where $C_{\pm }$ are the normalization factors. The superscripts of $f_{\pm
}^{\pm }$ and $\eta _{1,2}^{\pm }$ stand for $E_{\pm }^{0}$, and the
subscripts of $f_{\pm }^{\pm }$ for parity, respectively. The expressions
for $f_{\pm }^{\pm }$ and $\eta _{1,2}^{\pm }$ are given by
\begin{eqnarray}
f_{+}^{\pm }(z) &=&\left. \frac{\cosh (\lambda _{1}z)}{\cosh (\frac{\lambda
_{1}L}{2})}-\frac{\cosh (\lambda _{2}z)}{\cosh (\frac{\lambda _{2}L}{2})}%
\right\vert _{E=E_{\pm }^{0}},  \label{feta1} \\
f_{-}^{\pm }(z) &=&\left. \frac{\sinh (\lambda _{1}z)}{\sinh (\frac{\lambda
_{1}L}{2})}-\frac{\sinh (\lambda _{2}z)}{\sinh (\frac{\lambda _{2}L}{2})}%
\right\vert _{E=E_{\pm }^{0}}, \\
\eta _{1}^{\pm } &=&\left. \frac{(\lambda _{1})^{2}-(\lambda _{2})^{2}}{%
\lambda _{1}\coth (\frac{\lambda _{1}L}{2})-\lambda _{2}\coth (\frac{\lambda
_{2}L}{2})}\right\vert _{E=E_{\pm }^{0}}, \\
\eta _{2}^{\pm } &=&\left. \frac{(\lambda _{1})^{2}-(\lambda _{2})^{2}}{%
\lambda _{1}\tanh (\frac{\lambda _{1}L}{2})-\lambda _{2}\tanh (\frac{\lambda
_{2}L}{2})}\right\vert _{E=E_{\pm }^{0}}.  \label{feta2}
\end{eqnarray}%
The energy spectra and wavefunctions of the lower block $h(-A_{1})$ of $%
H_{0} $ can be obtained directly by replacing $A_{1}$ by $-A_{1}$. Based on
the above discussions, the four eigenstates of $H_{0}$ can be given by
\begin{eqnarray}
\Phi _{1} &=&\left[
\begin{array}{c}
\varphi (A_{1}) \\
0%
\end{array}%
\right] ,\ \Phi _{2}=\left[
\begin{array}{c}
\chi (A_{1}) \\
0%
\end{array}%
\right] ,  \nonumber  \label{H0states} \\
\Phi _{3} &=&\left[
\begin{array}{c}
0 \\
\varphi (-A_{1})%
\end{array}%
\right] ,\ \Phi _{4}=\left[
\begin{array}{c}
0 \\
\chi (-A_{1})%
\end{array}%
\right] ,
\end{eqnarray}%
with $\Phi _{1}\rightarrow \Phi _{3}$ and $\Phi _{2}\rightarrow \Phi _{4}$
under the time-reversal operation. We should emphasize that these four
solutions are for the surface states, and the solutions for the bulk states
are not presented here. We use the four states as the basis states, and
other states are discarded (except that in Fig. \ref{fig:wavefunction} where
four extra bulk states are also included by the same approach), because of a
large gap between the valence and conduction bands.

\subsection{Effective model for 3DTI films}

With the help of the four states Eq. (\ref{H0states}) at the $\Gamma $
point, we can expand Hamiltonian Eq. (\ref{H_zhang_natphys}) to obtain a new
effective Hamiltonian
\begin{equation}
H_{\mathrm{eff}}\equiv \int_{-L/2}^{L/2}dz[\Phi _{1},\Phi _{4},\Phi
_{2},\Phi _{3}]^{\dag }H[\Phi _{1},\Phi _{4},\Phi _{2},\Phi _{3}],
\end{equation}%
where for convenience, we organize the sequence of the basis states
following $\{\Phi _{1}$, $\Phi _{4}$, $\Phi _{2}$, $\Phi _{3}\}$. Under the
reorganized basis, the effective Hamiltonian is found as
\begin{equation}
H_{\mathrm{eff}}=\left[
\begin{array}{cc}
h_{+} & 0 \\
0 & h_{-}%
\end{array}%
\right] ,  \label{Heff2}
\end{equation}%
with
\begin{eqnarray}
h_{+} &=&E_{0}-Dk^{2}+\left[
\begin{array}{cc}
\frac{\Delta }{2}-Bk^{2} & \tilde{A}_{2}k_{-} \\
\tilde{A}_{2}^{\ast }k_{+} & -\frac{\Delta }{2}+Bk^{2}%
\end{array}%
\right] ,  \nonumber  \label{hphm} \\
h_{-} &=&E_{0}-Dk^{2}+\left[
\begin{array}{cc}
-\frac{\Delta }{2}+Bk^{2} & -\tilde{A}_{2}^{\ast }k_{-} \\
-\tilde{A}_{2}k_{+} & \frac{\Delta }{2}-Bk^{2}%
\end{array}%
\right] ,
\end{eqnarray}%
and
\begin{eqnarray}
B &=&\frac{\tilde{B}_{1}-\tilde{B}_{2}}{2},\ \ D=\frac{\tilde{B}_{1}+\tilde{B%
}_{2}}{2}-D_{2},  \nonumber  \label{paradefinition} \\
E_{0} &=&(E_{+}+E_{-})/2,\ \ \Delta =E_{+}-E_{-},  \nonumber \\
\tilde{B}_{1} &=&B_{2}\langle \varphi (A_{1})|\sigma _{z}|\varphi
(A_{1})\rangle ,  \nonumber \\
\tilde{B}_{2} &=&B_{2}\langle \chi (A_{1})|\sigma _{z}|\chi (A_{1})\rangle ,
\nonumber \\
\tilde{A}_{2} &=&A_{2}\langle \varphi (A_{1})|\sigma _{x}|\chi
(-A_{1})\rangle .
\end{eqnarray}%
We find that $\tilde{A}_{2}$ here can be either real or purely imaginary
(see Appendix \ref{sec:$AV$} for details), classifying the model into two
cases:

Case I is for a real $\tilde{A}_{2}\equiv \hbar v_{\mathrm{F}}$, and the
effective Hamiltonian is further written as
\begin{equation}
h_{\tau _{z}}=E_{0}-Dk^{2}+\hbar v_{\mathrm{F}}\tau _{z}\vec{\sigma}\cdot
\vec{k}+\tau _{z}\sigma _{z}(\frac{\Delta }{2}-Bk^{2}),\ \ \mathrm{(case\ I)}%
;  \label{hamiltonianbhz} \\
\end{equation}%
and case II for a purely imaginary $\tilde{A}_{2}\equiv i\hbar v_{\mathrm{F}%
} $,
\begin{equation}
h_{\tau _{z}}=E_{0}-Dk^{2}+\hbar v_{\mathrm{F}}(\vec{\sigma}\times \vec{k}%
)_{z}+\tau _{z}\sigma _{z}(\frac{\Delta }{2}-Bk^{2}),\ \ \mathrm{(case\ II)}
\label{hamilton2}
\end{equation}%
with $\tau _{z}=\pm 1$ corresponding to the upper (lower) $2\times 2$ block
in Eq. (\ref{Heff2}), $v_{\mathrm{F}}$ the defined Fermi velocity and $\vec{%
\sigma}$ and $\vec{k}$ here only refer to the components in $x$-$y$ plane.
In fact, these two effective Hamiltonian can consist of the invariants of
the irreducible representation $D_{1/2}$ of SU(2) group\cite{Winkler}.

Eq. (\ref{hphm}) can also be expressed in terms of the Pauli matrices
\begin{equation}
h_{\tau _{z}}=E_{0}-Dk^{2}+\mathbf{d}\cdot \mathbf{\sigma },  \label{h_pauli}
\end{equation}%
with
\begin{equation}
\mathbf{d}=\left\{
\begin{array}{ll}
\tau _{z}(\hbar v_{\mathrm{F}}k_{x},\ \hbar v_{\mathrm{F}}k_{y},\ \frac{%
\Delta }{2}-Bk^{2}), & \mathrm{(case\ I)} \\
&  \\
(\hbar v_{\mathrm{F}}k_{y},\ -\hbar v_{\mathrm{F}}k_{x},\ \tau _{z}(\frac{%
\Delta }{2}-Bk^{2})). & \mathrm{(case\ II)}%
\end{array}%
\right.   \label{hphm2}
\end{equation}%
$\mathbf{d}(k)$ vectors in case I and case II, respectively, correspond to
Dresselhaus- and Rashba-like textures. Note that case I is essentially the
effective 4$\times $4 model for the CdTe/HgTe quantum wells\cite%
{Bernevig2006.science.314.1757}. However, we find that case I only occurs
for quite a small range of thickness. For most thicknesses of interest, $%
\tilde{A}_{2}$ is pure imaginary. Therefore, we only focus on case II in the
following discussions. By far, we have reduced the anisotropic 3D Dirac
model into a generalized effective model for 2D thin films, under the
freestanding open boundary conditions.

\subsection{Effective continuous model for surface states}

Despite the simple explicit form, the parameters in Hamiltonian (\ref{Heff2}%
) need to be determined numerically. Before that, we can take two limits to
see their behaviors. The first limit is $\lambda _{\alpha }L\gg 1$, for $%
\alpha =1,2$. In this case, $\tanh (\frac{\lambda _{\alpha }L}{2})\backsimeq
1$, and both Eqs. (\ref{trans_eqn_1}) and (\ref{trans_eqn_2}) reduce to
\begin{equation}
(C-M-E-D_{+}\lambda _{1}^{2})\lambda _{2}=(C-M-E-D_{+}\lambda
_{2}^{2})\lambda _{1}.
\end{equation}
Solving this equation, we have an effective continuous model for the surface
states (ss) of 3D topological insulator as
\begin{eqnarray}
H_{\mathrm{ss}} &=&C+\frac{D_{1}M}{B_{1}}+(D_{2}-B_{2}\frac{D_{1}}{B_{1}}
)k^{2}  \nonumber \\
&&+A_{2}\sqrt{1-(\frac{D_{1}}{B_{1}})^{2}}(\sigma _{x}k_{y}-\sigma
_{y}k_{x}),
\end{eqnarray}
which has the same dispersion as Eq. (\ref{E_infinite}) and the same Fermi
velocity $v_{\mathrm{F}}=\frac{A_{2}}{\hbar }\sqrt{1-(\frac{D_{1}}{B_{1}}
)^{2}}$ as for the semi-infinite boundary conditions. In an isotropic case, $%
D_{1}=D_{2}$ and $B_{1}=B_{2}$, the quadratic term disappears and we have a
linear dispersion for the Dirac cone. Finally it is noticed that the models
for the surface states at the top and bottom surface have the same form
assumed $\lambda _{\alpha }L\gg 1.$ We will see that these results work well
even for films down to five quintuple layers (QL) of atoms in thickness (1
QL is about 1 nm).

\subsection{The ultra-thin limit}

Another opposite limit is $L\rightarrow 0$, which is a little bit
complicated since $\lambda _{\alpha }L$ does not approach to zero when $L$
is very small. In Eq. (\ref{trans_eqn_1}), the left side has an order of $%
L^{2}$ when $L\rightarrow 0$, so $\tanh (\frac{\lambda _{1}L}{2})$ must have
the order of $L^{-2}$, which means
\begin{equation}
\tanh (\frac{\lambda _{1}L}{2})=0\Rightarrow \lambda _{1}=i\frac{\pi }{L}.
\end{equation}
Combining this result with Eq. (\ref{lambda_alpha}), the model becomes
\begin{equation}  \label{ultrathin}
h_{\tau _{z}}=\frac{D_{1}\pi ^{2}}{L^{2}}+D_{2}k^{2}+A_{2}(\vec{\sigma}
\times \vec{k})_{z}+\tau _{z}(\frac{B_{1}\pi ^{2}}{L^{2}}+B_{2}k^{2})\sigma
_{z}.
\end{equation}
It is found that a finite energy gap opens at $k=0$, i.e., $%
\Delta=2B_1\pi^2/L^2$ as shown in Fig. \ref{fig:gap}. Note that this result in the $L\rightarrow 0$ limit even provides a rough estimate of the gap for most thicknesses. Besides, the continuum limit generally assumed in this work also works well even for only several quintuple layers.

\section{\label{sec:SIA} Structure Inversion Asymmetry}

\subsection{Structure Inversion Asymmetry}

A recent experiment\cite{Zhang2009.arXiv.0911.3706} revealed that the
substrate on which the film is grown influences dramatically electronic
structure inside the film. Because the top surface of the film is usually
exposed to the vacuum and the bottom surface is attached to a substrate, the
inversion symmetry does not hold along $z$ direction, leading to the
Rashba-like energy spectra for the gapped surface states. In this case, an
extra term that describes the structure inversion asymmetry (SIA) needs to
be taken into account in the effective model.

We use the same method as that in Sec. \ref{sec:perturbative} to include the
SIA term. Without loss of generality, we add a potential energy $V(z)$ into
the Hamiltonian. Generally speaking, $V(z)$ can be expressed as $%
V(z)=V_{s}(z)+V_{a}(z)$, in which $V_{s}(z)=V_{s}(-z)$ and $%
V_{a}(z)=-V_{a}(-z)$. The symmetric term $V_{s}$ could contribute to the
mass term $\Delta $ in the effective model, which may lead to an energy
splitting of the Dirac cone at the $\Gamma $ point. We do not discuss it in
details in this paper. Here we focus on the case of the antisymmetric term, $%
V(z)=V_{a}(z)$, which breaks the top-bottom inversion symmetry in the
Hamiltonian. A detailed analysis demonstrates that $V_{a}(z)$ couples $\Phi
_{1}$ $(\Phi _{3})$ to $\Phi _{2}$ $(\Phi _{4})$, which can be readily seen
according to their spin and parity natures. The modified effective
Hamiltonian in the presence of $V(z)$ becomes
\begin{equation}
H_{\mathrm{eff}}^{\mathrm{SIA}}=H_{\mathrm{eff}}+\left[
\begin{array}{cccc}
0 & 0 & \tilde{V} & 0 \\
0 & 0 & 0 & \tilde{V}^{\ast } \\
\tilde{V}^{\ast } & 0 & 0 & 0 \\
0 & \tilde{V} & 0 & 0%
\end{array}%
\right] ,  \label{H_SIA}
\end{equation}%
where
\begin{equation}
\tilde{V}=\int_{-L/2}^{L/2}dz\langle \varphi (A_{1})|V_{a}(z)|\chi
(A_{1})\rangle .  \label{V_def}
\end{equation}%
Comparing this definition with that of $\tilde{A}_{2}$ in Eq. (\ref%
{paradefinition}), we find that $\tilde{V}$ also can be either real or
purely imaginary. In the case of a purely imaginary (case II) $\tilde{A}_{2}$%
, $\tilde{V}$ must be real (see Appendix \ref{sec:$AV$}), and the effective
Hamiltonian with SIA can be written as
\begin{equation}
H_{\mathrm{eff}}^{\mathrm{SIA}}=\left[
\begin{array}{cc}
h_{+}(k) & \tilde{V}\sigma _{0}\nonumber \\
\tilde{V}\sigma _{0} & h_{-}(k)\nonumber%
\end{array}%
\right] .
\end{equation}%
In the case of a real $\tilde{A}_{2}$, $\tilde{V}$ must be purely imaginary,
and the effective Hamiltonian with SIA then has the form
\begin{equation}
H_{\mathrm{eff}}^{\mathrm{SIA}}=\left[
\begin{array}{cc}
h_{+}(k) & \tilde{V}\sigma _{z}\nonumber \\
-\tilde{V}\sigma _{z} & h_{-}(k)\nonumber%
\end{array}%
\right] .
\end{equation}

\begin{figure}[tbph]
\centering \includegraphics[width=0.5\textwidth]{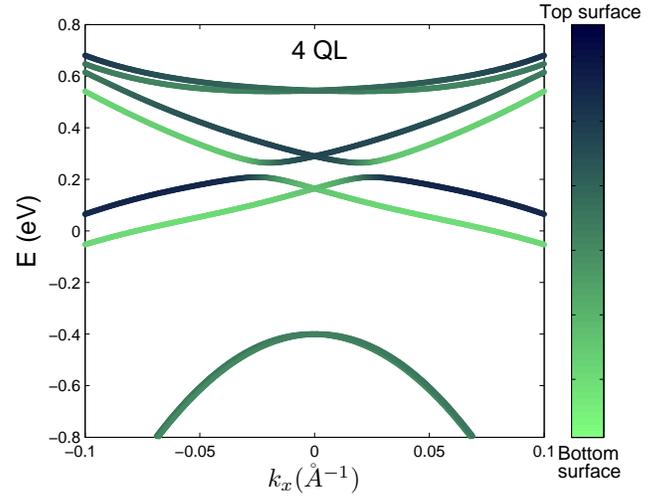}
\caption{(Color online) Energy spectra of surface states (four branches in
the middle) and several branches of bulk states (those at the top and the
bottom) for a film with the thickness of 4 QL in the presence of the
structure inversion asymmetry. The color of lines corresponds to the spatial
distribution of the wavefunctions in $z$ direction. Dark blue (light green)
represents that the wavefunctions mainly distribute on the side of the top
(substrate) surface. The model parameters are listed in the first row of
Table \protect\ref{tab:4QLDiracParameters}. }
\label{fig:wavefunction}
\end{figure}

\begin{figure}[htbp]
\centering \includegraphics[width=0.4\textwidth]{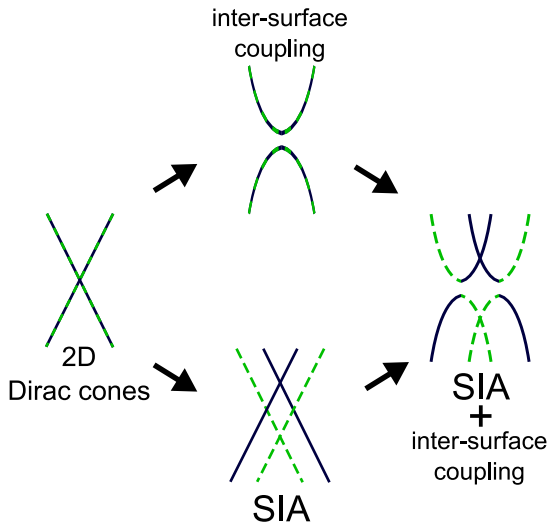}
\caption{(Color online) The evolution of the doubly-degenerate gapless Dirac
cones for the 2D surface states, in the presence of both the inter-surface
coupling and structure inversion asymmetry, into the gapped hyperbolas that
also split in $k$-direction. The blue solid and green dashed lines
correspond to the states residing near the top and bottom surfaces,
respectively.}
\label{fig:SIAGap}
\end{figure}

Without the SIA term, the effective Hamiltonian (\ref{hamilton2}) gives the
energy spectra of the gapped surface states as
\begin{equation}
E_{\pm }=E_{0}-Dk^{2}\pm \sqrt{(\frac{\Delta }{2}-Bk^{2})^{2}+(\hbar v_{%
\mathrm{F}})^{2}k^{2}},  \label{spin}
\end{equation}%
where $+$ $(-)$ sign stands for the conduction (valence) band, each of which
has double spin degeneracy due to time-reversal symmetry. When the SIA term
is included, the Hamiltonian (\ref{H_SIA}) gives
\begin{eqnarray}
E_{1,\pm } &=&E_{0}-Dk^{2}\pm \sqrt{(\frac{\Delta }{2}-Bk^{2})^{2}+(|\tilde{V%
}|+\hbar v_{\mathrm{F}}k)^{2}},  \nonumber  \label{E_SIA} \\
E_{2,\pm } &=&E_{0}-Dk^{2}\pm \sqrt{(\frac{\Delta }{2}-Bk^{2})^{2}+(|\tilde{V%
}|-\hbar v_{\mathrm{F}}k)^{2}},  \nonumber \\
&&
\end{eqnarray}%
where the extra index 1 (2) stands for the inner (outer) branches of the
conduction or valence bands. The energy spectra in the presence of $\tilde{V}
$ is shown in Fig. \ref{fig:wavefunction}. Each spin-degenerate dispersion
in Eq. (\ref{spin}) shifts away from each other along $k$ axis. Both the
conduction and valence bands show Rashba-like splitting. An intuitive
understanding of the energy spectra in Fig. \ref{fig:wavefunction} can be
given with the help of Fig. \ref{fig:SIAGap}. On the left is for a thicker
freestanding symmetric TI film, and it has a single gapless Dirac cone on
each of its two surfaces, with the solid and dash lines for the top and
bottom surface, respectively. The two Dirac cones are degenerate. The top of
Fig. \ref{fig:SIAGap} indicates that the inter-surface coupling across an
ultrathin film will turn the Dirac cones into gapped Dirac hyperbolas. On
the bottom of Fig. \ref{fig:SIAGap}, SIA lifts the Dirac cone at the top
surface while lowers the Dirac cone at the bottom surface. The potential
difference at the top and bottom surfaces removes the degeneracy of the
Dirac cones. On the right of Fig. \ref{fig:SIAGap}, the coexistence of both
the inter-surface coupling and SIA leads to two gapped Dirac hyperbolas
which also split in $k$-direction, as shown in Fig. \ref{fig:wavefunction}.

\subsection{Location of the surface states}

Location of the surface states can be revealed by evaluating the expectation
of position z of these states. The spatial distributions along the z
direction of a state $\psi _{\alpha }$ can be evaluated by the expectation
of position in $z$ direction $\langle z\rangle $,
\begin{equation}
\langle z\rangle _{\alpha }=\int_{-\frac{L}{2}}^{\frac{L}{2}}z|\psi _{\alpha
}|^{2}dz.
\end{equation}
By this definition, $\langle z\rangle _{\alpha }\in \lbrack -\frac{L}{2},
\frac{L}{2}]$ and $\langle z\rangle _{\alpha }$ becomes $0$ for a symmetric
spatial distribution.

With the SIA, the eigen wavefunctions are found as
\begin{equation}  \label{psi1pm}
\psi _{1\pm }=\frac{1}{2\sqrt{E_{\pm }^{\mathrm{in}}(E_{\pm }^{\mathrm{in}
}+t)}k}\left[
\begin{array}{c}
i(t+E_{\pm }^{\mathrm{in}})k \\
(|\tilde{V}|+\hbar v_{\mathrm{F}}k)k_{+} \\
i(|\tilde{V}|+\hbar v_{\mathrm{F}}k)k \\
(t+E_{\pm }^{\mathrm{in}})k_{+}%
\end{array}
\right] ,
\end{equation}
with $E_{\pm }^{\mathrm{in}}=E_{1\pm }-E_{0}+Dk^{2}$, $t=\frac{\Delta }{2}
-Bk^{2}$, and,
\begin{equation}  \label{psi2pm}
\psi _{2\pm }=\frac{1}{2\sqrt{E_{\pm }^{\mathrm{out}}(E_{\pm }^{\mathrm{out}
}+t)}k}\left[
\begin{array}{c}
-i(t+E_{\pm }^{\mathrm{out}})k \\
(|\tilde{V}|-\hbar v_{\mathrm{F}}k)k_{+} \\
i(-|\tilde{V}|+\hbar v_{\mathrm{F}}k)k \\
(t+E_{\pm }^{\mathrm{out}})k_{+}%
\end{array}
\right]
\end{equation}
with $E_{\pm }^{\mathrm{out}}=E_{2\pm }-E_{0}+Dk^{2}$. Fig. \ref%
{fig:wavefunction} demonstrates $\langle z\rangle $ by the brightness of
lines, with dark blue for $\langle z\rangle =\frac{L}{2}$ (the top surface),
and light green for $\langle z\rangle =-\frac{L}{2}$ (the substrate or
bottom surface).

For a thin film of 4 quintuple layers (QL), $L=3.8$nm, it is found that the
two surface states are well separated and dominantly distributed near the
two surfaces. The averaged $\langle z\rangle \simeq \pm \frac{L}{3}$, which
is about 2/3 of a QL ($\approx L/6$) deviating from the surface. In this
case the top and bottom surface states are well defined even without the SIA
($\tilde{V}=0$). The average value remains almost unchanged in a large range
of $k$. However, the crossing point of the spectra of the top and bottom
surface states, the averaged $\langle z\rangle $ changes from $+L/3$ to 0$,$
and then goes to the value of $-L/3$. This demonstrates that the finite
thickness makes the two states couple with each other as their wave
functions along the z direction have a finite overlap. As a result the two
states open an energy gap as in the case of edge states in QSH system\cite%
{Zhou2008.prl.101.246807}. The value of the gap is a function of $L$ as
shown in Fig. 2(a) and (b). Near this region, $\langle z\rangle $ varies
from $\langle z\rangle \simeq L/3$ to $-L/3$, and becomes zero exactly when
two states are mixed completely. For a large $L$, we find that the averaged
distance of the surface states deviating from the surface remains about $%
1$ QL.

Simply speaking, the states close to the top surface are easier to be probed
by light than those close to the bottom surface. This provides a hint to
understand why there are branches in energy spectra with much more faint
ARPES signals\cite{Zhang2009.arXiv.0911.3706}.

\section{\label{sec:QSH} Thin film B$\mathrm{i}_{2}$S$\mathrm{e}_{3}$ and
QSH states}

In this section, we will investigate the realization of QSH effect in thin
films and apply the effective model to the thin film Bi$_{2}$Se$_{3}$. When
the system does not break the inversion symmetry, the effective Hamiltonian
is block-diagonalized by $\tau _{z}=\pm 1$. This is in a good agreement with
the theory by Murakami et al\cite{Murakami2007.prb.76.205304}. In this case
we can define a $\tau _{z}$ -dependent Chern number (Hall conductance) for
each block like the spin Chern number\cite{Sheng.prl.97.036808}, from which
the nontrivial QSH phase can be identified. After introducing the SIA term,
the $\tau _{z}$-dependent Chern number loses its meaning as the two blocks
are mixed together. However, we can still employ the Z$_{2}$ topological
classification\cite{Kane2005.prl.95.146802}, which requires no inversion
symmetry, to identify possible QSH thin films in experiment.

\subsection{QSH effect without SIA}

Considering the block-diagonal form of the effective model without SIA (\ref%
{Heff2}), we can derive the Hall conductance for each block, separately. For
the $2\times 2$ Hamiltonian in terms of the $\mathbf{d}(k)$ vectors and
Pauli matrices in Eq. (\ref{h_pauli}), the Kubo formula for the Hall
conductance can be generally expressed as \cite%
{Qi2006.prb.74.085308,Zhou2006.prb.73.165303}
\begin{equation}  \label{hallformula}
\sigma _{xy}=-\frac{e^{2}}{2\Omega \hbar }\sum_{k}\frac{(f_{k,-}-f_{k,+})}{
d^{3}}\epsilon _{\alpha \beta \gamma }\frac{\partial d_{a}}{\partial k_{x}}
\frac{\partial d_{\beta }}{\partial k_{y}}d_{\gamma }
\end{equation}
where $\Omega $ is the volume of the system, $d$ the norm of $%
(d_{x},d_{y},d_{z})$, $f_{k,\pm }=$ $1/\{\exp [(E_{\pm }(k)-\mu
)/k_{B}T]+1\} $ the Fermi distribution function of electron ($+$) and hole ($%
-$) bands, with $\mu $ the chemical potential, $k_{B}$ the Boltzmann
constant, and $T$ the temperature.

At zero temperature and when the chemical potential $\mu $ lies between the
band gap $(-\frac{|\Delta |}{2},\frac{|\Delta |}{2})$, the Fermi functions
reduce to $f_{k,+}=0$ and $f_{k,-}=1$. In this case we have\cite%
{Lu2009.arxiv}
\begin{equation}
\sigma _{xy}^{\tau _{z}}=-\tau _{z}\frac{e^{2}}{2h}[\mathrm{sgn}(\Delta )+%
\mathrm{sgn}(B)].  \label{sigmaxy}
\end{equation}%
This result intuitively shows that only when $B$ and $\Delta $ have the same
sign, the Chern number is equal to $+1$ or $-1$, which is topologically
nontrivial, and the Hall conductance is quantized to be $\pm e^{2}/h$. In
other words, the QSH depends not only on the sign of $\Delta $ at the $%
\Gamma $ point but also on that of $B$ for $k$ large enough. Experimentally,
the $\tau _{z}$-dependent Hall conductance can be probed by the nonlocal
measurement, just like that for the 2D CdTe/HgTe quantum wells\cite%
{Roth2009.science.325.294}.

\subsection{QSH effect with SIA: Z$_{2}$ invariant}

\begin{figure}[htbp]
\centering \includegraphics[width=0.5\textwidth]{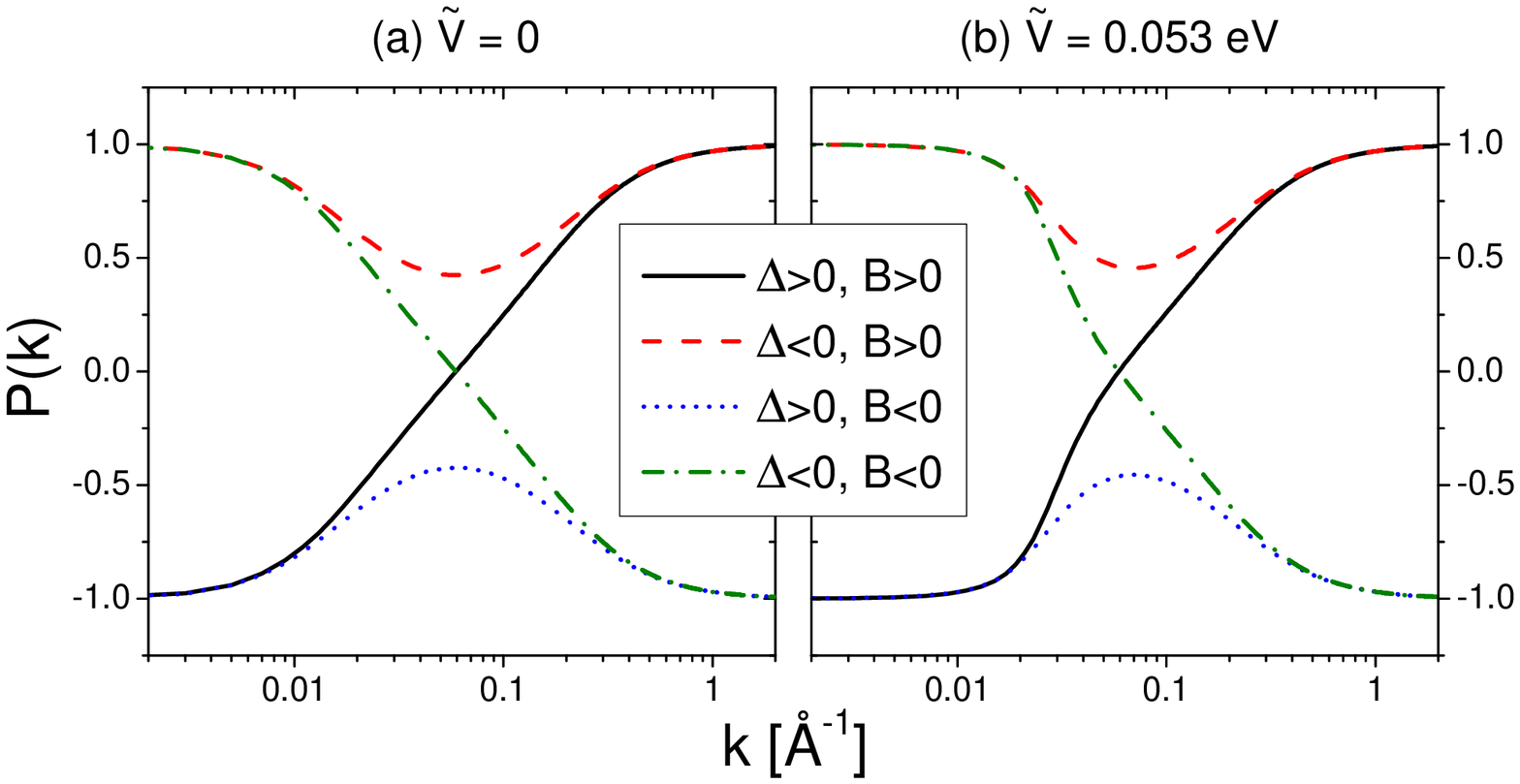} %
\includegraphics[width=0.4\textwidth]{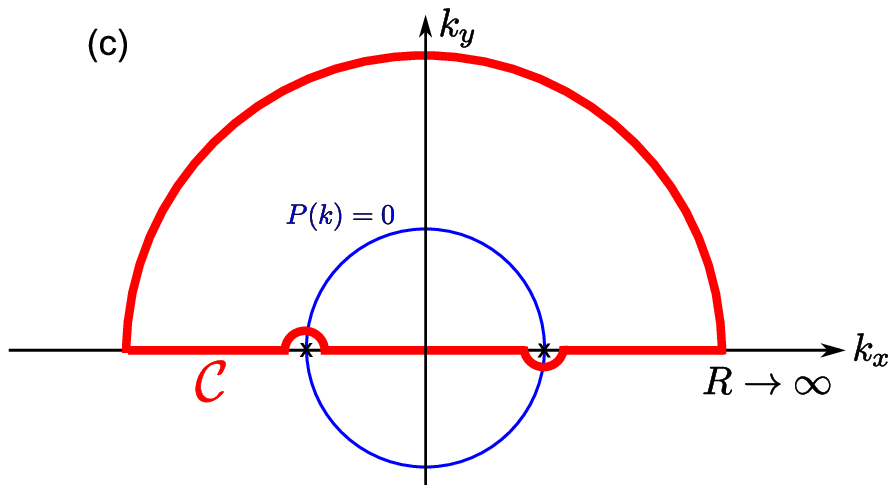}
\caption{ (Color online) [(a) and (b)] $P(k)$ for four combinations of $%
\Delta$ and $B$, in absence (a) and the presence (b) of a small SIA $\tilde{%
V }$. $|\Delta|$=0.07 eV and $|B|$=10 eV \AA $^2$. (c) The contour $\mathcal{%
C} $ is used to count the number of the pairs of the zeros of $P(k)$, which
form a circle ring when $\Delta B>0$.}
\label{fig:pfaffian}
\end{figure}

In the presence of SIA, $\tilde{V}$ couples the blocks $h_{+}$ and $h_{-}$,
so the $\tau _{z}$-dependent Hall conductance becomes nonsense. Following
Kane and Mele\cite{Kane2005.prl.95.146802}, we can employ the Z$_{2}$
topological classification to give a criterion of the QSH phase, because it
does not require inversion symmetry as a necessary condition. The Z$_{2}$
index can be obtained by counting the number of pairs of complex zeros of $P(%
\mathbf{k})\equiv \mathrm{Pf}[A(\mathbf{k})]$, where the \emph{Pfaffian} is
defined as
\begin{equation}
\mathrm{Pf}[A(\mathbf{k})]=\frac{1}{2^{n}n!}\sum_{\mathrm{\ Permutations}%
\newline
\mathrm{of}\
\{i_{1},...,i_{2n}\}}(-1)^{N}A_{i_{1}i_{2}}...A_{i_{2n-1}i_{2n}},
\end{equation}%
in which $N$ counts the number of times of permutations, and $A(\mathbf{k})$
is a $2n$ order anti-symmetric matrix defined by the overlaps of time
reversal
\begin{equation}
A_{ij}(\mathbf{k})=\langle u_{i}(\mathbf{k})|\Theta |u_{j}(\mathbf{k})\rangle
\end{equation}%
with $i,j$ run over all the bands below the Fermi surface, i.e., $\psi _{1-}$
and $\psi _{2-}$ in the present case according to Eqs. (\ref{psi1pm}) and (%
\ref{psi2pm}). Based on the spin nature of the basis states $\{\Phi
_{1},\Phi _{4},\Phi _{2},\Phi _{3}\}$ in our effective model, the
time-reversal operator here is defined as $\Theta \equiv i\sigma _{x}\otimes
\sigma _{y}\mathcal{K}$, where $\sigma _{x}$ and $\sigma _{y}$ are the $x$-
and $y$-component of Pauli matrices, respectively, and $\mathcal{K}$ the
complex conjugate operator. The number of pairs of zeros can be counted by
evaluating the winding of the phase of $P(\mathbf{k})$ around a contour $C$
enclosing half of the complex plane of $\mathbf{k}=k_{x}+ik_{y}$,
\begin{equation}
I=\frac{1}{2\pi i}\oint_{C}d\mathbf{k}\cdot \nabla _{\mathbf{k}}\mathrm{log}%
[P(\mathbf{k})+i\delta ].
\end{equation}%
Because the model is isotropic, we can choose $\mathcal{C}$ to enclose the
upper half plane, the integral then reduces to only the path along $k_{x}$
-axis while the part of the half-circle integral vanishes for $\delta >0$
and $\left\vert \mathbf{k}\right\vert \rightarrow +\infty $.

In the absence of the SIA term, $P(\mathbf{k})$ is found for the Hamiltonian
(\ref{hamilton2}) as
\begin{equation}  \label{PknoV}
P(k)=\frac{-\frac{\Delta }{2}+Bk^{2}}{\sqrt{(\frac{\Delta }{2}
-Bk^{2})^{2}+(\hbar v_{\mathrm{F}})^{2}k^{2}}},
\end{equation}
in which one can check that the zero points exist only when $k^{2}=\Delta
/2B>0$, and form a circle ring. Along $k_{x}$-axis only one of a pair of
zeros in the ring is enclosed in the contour $\mathcal{C}$, which gives a $%
Z_{2}$ index $I=1$. This defines the nontrivial QSH phase, and is in
consistence with the conclusion by the Hall conductance in Eq. (\ref{sigmaxy}%
).

In the presence of a small SIA term $\tilde{V}<\hbar v_{\mathrm{F}}\sqrt{
|\Delta /2B|}$, with the help of the eigen wavefunctions (\ref{psi1pm}) and (%
\ref{psi2pm}), real $P(\mathbf{k})$ can be found (after a $U(1)$ rotation)
as
\begin{eqnarray}
P(k) &=&\frac{(t+E_{-}^{\mathrm{in}})(t+E_{-}^{\mathrm{out}})+|\tilde{V}
|^{2}-(\hbar v_{\mathrm{F}}k)^{2}}{2\sqrt{E_{-}^{\mathrm{in}}E_{-}^{\mathrm{%
\ out}}(t+E_{ -}^{\mathrm{in}})(t+E_{-}^{\mathrm{out}})}}  \nonumber
\label{Pk_small_V} \\
&&\times \left\{
\begin{array}{cc}
\mathrm{sgn}(\hbar v_{\mathrm{F}}k-|\tilde{V}|), & \Delta >0 \\
1, & \Delta \leq 0%
\end{array}
\right. ,
\end{eqnarray}
where the $\mathrm{sgn}$ is to secure the continuity of $P(\mathbf{k})$. One
can check that $P(0)=-\mathrm{sgn}(\Delta )$ and $P(\infty )=\mathrm{sgn}(B)$
. Besides, for a small $\tilde{V}$, the behavior of $P(\mathbf{k})$ between $%
P(0)$ and $P(\infty )$ will not change qualitatively (see Fig. \ref%
{fig:pfaffian}). Therefore, for $\Delta B>0$, $P(k_{x},0)$ should still have
odd pairs of zeros. For a large $\tilde{V}\geq \hbar v_{\mathrm{F}}\sqrt{
|\Delta /2B|}$,
\begin{eqnarray}  \label{Pk_large_V}
P(k) &=&\frac{(t+E_{-}^{\mathrm{in}})(t+E_{-}^{\mathrm{out}})+|\tilde{V}
|^{2}-(\hbar v_{\mathrm{F}}k)^{2}}{2\sqrt{E_{-}^{\mathrm{in}}E_{-}^{\mathrm{%
\ out}}(t+E_{-}^{\mathrm{in}})(t+E_{-}^{\mathrm{out}})}}  \nonumber \\
&&\times \left\{
\begin{array}{ccc}
\mathrm{sgn}(\hbar v_{\mathrm{F}}k-|\tilde{V}|), & B<0 &  \\
1, & B\geq 0 &
\end{array}
\right. .
\end{eqnarray}
One can check for this case $P(0)P(\infty )$ is always positive thus $P(k)$
has even pairs of zeros, regardless of the signs and values of $\Delta $ and
$B$. In other words, a large SIA will always destroy the quantum spin Hall
phase.

\subsection{Thin film Bi$_{2}$Se$_{3}$ and QSH effect}

\begin{table}[tbph]
\caption{Fitting parameters to the Bi$_{2}$Se$_{3}$ thin films, using the
energy spectra Eq. (\protect\ref{E_SIA}) from our effective model.[adopted
from Ref. \protect\cite{Zhang2009.arXiv.0911.3706}]}
\label{tab:thickness}
\begin{ruledtabular}
\begin{tabular}{ccccccc}
Layers & $E_0$ & $D$ & $\Delta$ & $B$ & $v_{\mathrm{F}}$ & $|\tilde{V}|$\\
(QL) & (eV) & (eV\AA $^2$) & (eV) & (eV\AA $^2$) & ($10^5$m/s) & (eV)\\ \hline
2 & -0.470 & -14.4 & 0.252 & 21.8 & 4.47 & 0 \\ \hline
3 & -0.407 & -9.7 & 0.138 & 18.0 & 4.58 & 0.038  \\ \hline
4 & -0.363 & -8.0 & 0.070 & 10.0 & 4.25 & 0.053  \\ \hline
5 & -0.345 & -15.3 & 0.041 & 5.0 & 4.30 & 0.057  \\ \hline
6 & -0.324 & -13.0 & 0 & 0 & 4.28 & 0.068  \\ 
\end{tabular}
\end{ruledtabular}
\end{table}

Recently, thickness-dependent band structure of molecular beam epitaxy grown
ultrathin films Bi$_{2}$Se$_{3}$ was investigated by in-situ angle-resolved
photoemission spectroscopy\cite{Zhang2009.arXiv.0911.3706}. An energy gap
was first observed experimentally in the surface states of Bi$_{2}$Se$_{3}$
below the thickness of six quintuple layers, which confirms theoretical
prediction as a finite size effect\cite{Zhou2008.prl.101.246807,Lu2009.arxiv,Liu2009.arxiv,Linder2009.prb.80.205401}%
.

Table \ref{tab:thickness} shows the fitting parameters to the ARPES data of
Bi$_{2}$Se$_{3}$ thin films\cite{Zhang2009.arXiv.0911.3706} using the energy
spectra formula [Eq. (\ref{E_SIA})]. For the films with thickness ranging
from 2 QL to 5 QL, all of them satisfy $\mathrm{sgn}(\Delta B)>0$ and $%
\tilde{V}<\hbar v_{\mathrm{F}}\sqrt{|\Delta /2B|}$, hence the films are
possibly in the QSH regime. We identify that only 2 QL, 3 QL, and 4 QL
belong to the nontrivial case for potential QSH effect. 5 QL is an
exceptional case that the fitted parameters $B$ and $D$ do not satisfy the
existence condition of an edge states solution\cite{Zhou2008.prl.101.246807}%
. The condition of $B^{2}<D^{2}$ will lead to the band gap closing for a
large $k$. However, it is understood that the model is only valid near the $%
\Gamma $ point, and the fitting parameters are limited to the case of small $%
k$. And the band gap was measured clearly for the film of 5QL.

It was previously predicted, using the parameters from the first-principles
calculation\cite{Zhang2009.natphys.5.438}, that the gap $\Delta $ should
oscillate as a function of the film thickness\cite%
{Lu2009.arxiv,Liu2009.arxiv,Linder2009.prb.80.205401}. However, this
oscillation is not reflected in the measured results.

\subsection{QSH effect of SIA and the edge states}

In the quantum Hall effect the Chern number of the bulk states has an
explicit correspondence to the number of edge states in an open boundary
condition\cite{Hatsugai1993.prl.71.3697}. In topological insulator or QSH
system, the Z$_{2}$ topological invariant has also a relation to the number
of helical edge states\cite{Qi2006.prb.74.045125}. As a supplementary
support to the above conclusion, we demonstrated the presence of edge states
in a periodic boundary condition along the x-direction and an open boundary
condition (say along y-direction) imposed in a geometry of strip of the thin
film by means of numerical calculation. Using the parameters in Table I, we
have concluded that a stripe of 2 - 4 QL will exhibit helical edge states.
More specifically, we present the energy dispersion for 4QL in Fig. \ref%
{fig:dispersion}. There is a doubly-degenerate Dirac point inside the gap of
the 2D surface states for 4 QL in consistence with the results obtained in
the above sections.

\begin{figure}[htbp]
\centering \includegraphics[width=0.5\textwidth]{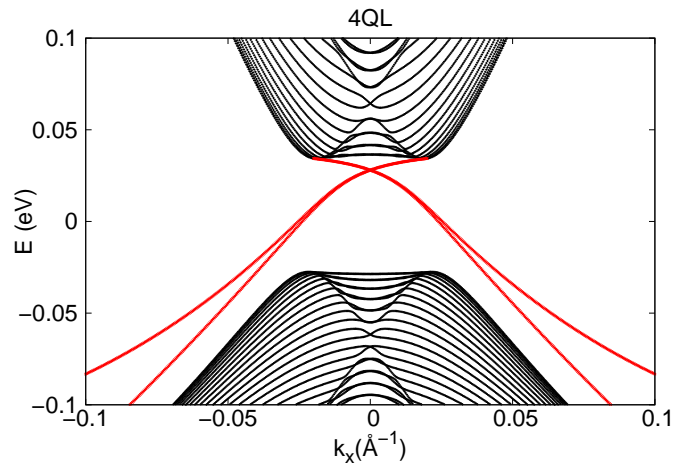}
\caption{ (Color online) The energy spectra by the tight binding calculation
for 4 QL of Bi$_2$Se$_3$ thin films with the width along $y$-direction 200
nm. The parameters are given in table \protect\ref{tab:thickness}.}
\label{fig:dispersion}
\end{figure}

\section{\label{sec:conclusion}Conclusions}

We derived two-dimensional effective continuous models for the surface
states and thin films of three-dimensional topological insulators (3DTI). A
gapless Dirac cone was confirmed for the surface states of a 3DTI. For a
thin film, the coupling between opposite topological surface states in space
opens an energy gap, and the Dirac cone evolves into a gapped Dirac
hyperbola. The thin film may break the top-bottom symmetry. For example, the
thin film grows on a substrate, and possesses the structural inversion
asymmetry (SIA). This SIA leads to a Rashba-like coupling and energy
splitting in the momentum space. It also leads to asymmetric distributions
of states along film growth direction.

The ARPES measurements on Bi$_{2}$Se$_{3}$ films have demonstrated that the
surface spectra opens a visible energy gap when the thickness is below 6QLs.%
\cite{Zhang2009.arXiv.0911.3706} The energy gap was observed to be a
function of the thickness of thin film, and in a good consistence with
theoretical prediction as a finite size effect of the thickness of thin
film. The Rashba-like splitting was measured clearly in the thin film of 2
to 6 QLs. This can be explained very well from the inclusion of the SIA.
Since the thin film was grown on a SiC substrate and the other surface is
exposed to the vacuum this fact results in the SIA in the thin film. Another
direct evidence to support the SIA is the signal intensity pattern of the
energy spectra of ARPES. Usually the surface states are located dominantly
near the top and bottom surfaces. The signal intensity for these two
branches of energy spectra of ARPES are different. The SIA will cause the
coupling between two surface states near their crossing point. That is why
the Rashba-like splitting of the ARPES spectra has a bright crossing point
near the $\Gamma $ point, with one branch bright and the other almost
invisible. Thus the SIA term can be used to describe the ARPES measurements
on the thin film Bi$_{2}$Se$_{3}$ very well.

Our effective model demonstrates that the 3DTI can be reduced to an
two-dimensional quantum spin Hall system due to the spatial confinement.
Strictly speaking, the system is no longer a 3DTI in the original sense once
the energy gap opens in the surface bands, since the Z2 invariant for the
bulk states becomes zero. However the surface bands themselves may
contribute a non-trivial one in the Z2 invariant even when the SIA term is
included. Our calculation demonstrates that a strong SIA always intends to
destroy the quantum spin Hall effect. A critical value for SIA exists, at
the point there is a transition from a topological trivial to non-trivial
phases. Based on the model parameters fitted from the experimental data of
ARPES, we conclude that the thin film Bi$_{2}$Se$_{3}$ should exhibit
quantum spin Hall effect once the energy gap opens in the surface spectra
due to the spatial confinement of the thin film.

\section*{Acknowledgments}

We thank Ke He and Qi-Kun Xue for providing experimental data prior to
publication, and Qian Niu for helpful discussions. This work was supported
by the Research Grant Council of Hong Kong under Grant No. HKU 7037/08P and
HKU 10/CRF/08.

\section{\label{sec:$AV$} Appendix: Model parameters $\tilde{A}_{2}$ and $%
\tilde{V}$}

\begin{table}[htpb]
\caption{Four possible combinations of $\protect\lambda_{1}$ and $\protect%
\lambda_{2}$, according to Eq. (\protect\ref{lambda_alpha}), and resulting $%
f_{\pm }$ and $\protect\eta_{1,2}$ according to Eq. (\protect\ref{H0states}%
). According to Eq. (\protect\ref{lambda_alpha}), $\protect\lambda_{1}^{2}<%
\protect\lambda_{2}^{2}$, so there does not exist a case when $\protect%
\lambda_{1}$ is real and $\protect\lambda_{2}$ is pure imaginary.}%
\begin{ruledtabular}
\begin{tabular}{c|cc|cc}
 &  $\lambda_1$&  $\lambda_2$ & $f_{\pm}$ & $\eta_{1,2}$ \\
  \hline
case A &  $\lambda_2^*$ & $\lambda_1^*$   & imaginary  & real \\
\hline
 &  real & real   & real  & real \\
case B &  imaginary & real   & real  & real \\
 &  imaginary & imaginary   & real  & real \\
\end{tabular}
\end{ruledtabular}
\end{table}

\begin{table}[htpb]
\caption{Four possible groups of $E_{+}$ and $E_{-}$, and resulting values
of $\tilde{A}_{2}$ and $\tilde{V}$.}
\label{tab:AV}%
\begin{ruledtabular}
\begin{tabular}{cc|cc}
   $E_+$&  $E_-$ & $\widetilde{A}_2$ & $\widetilde{V}$ \\
  \hline
   case A & case A   & imaginary  & real \\
   case A & case B   & real  & imaginary \\
   case B & case A   & real  & imaginary \\
   case B & case B   & imaginary  & real \\
\end{tabular}
\end{ruledtabular}
\end{table}

In this appendix, we demonstrate that both the parameters
$\tilde{A}_{2}$ and $\tilde{V}$ in the effective model (\ref{Heff2})
can be either real or purely imaginary, and the product of
$\tilde{A}_{2}\tilde{V}$ must be pure
imaginary. By putting the wavefunctions Eqs. (\ref{phichi1}) and (\ref%
{phichi2}) into the definitions in Eqs. (\ref{paradefinition}) and (\ref%
{V_def}), we have

\begin{eqnarray}  \label{definition of A}
\tilde{A}_{2} &=&iA_{1}A_{2}D_{+}C_{+}C_{-}\int_{-\frac{L}{2}}^{\frac{L}{2}
}dz[\eta _{2}^{-}f_{+}^{+\ast }f_{+}^{-}+\eta _{1}^{+\ast }f_{-}^{+\ast
}f_{-}^{-}]  \nonumber \\
\tilde{V} &=&C_{+}C_{-}\int_{-\frac{L}{2}}^{\frac{L}{2}}dzV(z)[D_{+}^{2}\eta
_{1}^{+\ast }\eta _{2}^{-}f_{-}^{+\ast }f_{+}^{-}+A_{1}^{2}f_{+}^{+\ast
}f_{-}^{-}].  \nonumber \\
\end{eqnarray}
For arbitrary energy, Eq. (\ref{lambda_alpha}) requires that the values of $%
\lambda _{1}$ and $\lambda_{2}$ can only be one of the combinations shown in
Tab. III. By putting these combinations into Eq. (\ref{feta1})-(\ref{feta2}%
), one can show that for the first entry, $\eta _{1}$ and $\eta _{2}$ are
real while $f_{+}$ and $f_{-}$ are pure imaginary, referred as the case A;
while for entries 2 - 4, all of $\eta _{1}$, $\eta _{2}$, $f_{+}$, and $f_{-}
$ are real, referred as the case B. For an arbitrary group of $E_{+}$ and $%
E_{-}$, each of them belongs to either the case A or B, leading to four
possibilities, as shown in Tab. \ref{tab:AV}. In particular, according to
Eq. (\ref{complexlambda12}), when $A_{1}^{2}/(-D_{+}D_{-})>4M/B_{1}$, there
is no complex $\lambda _{1,2}$, corresponding to the last row of Tab. \ref%
{tab:AV}, i.e., $\tilde{A}_{2}$ is pure imaginary while $\tilde{V}$ is real.

\end{document}